\documentclass[10pt]{article}
\usepackage[OE]{express}
\usepackage{tabularx}
\usepackage{float}
\usepackage{amsmath}
\usepackage{nicefrac}
\usepackage{lipsum}
\usepackage{multicol}
\usepackage{multirow}
\usepackage{graphicx}
\usepackage{textcomp}

\begin{document}
\title{Integrating quantum key distribution with classical communications in backbone fiber network}

\author{Yingqiu Mao,\authormark{1,2} Bi-Xiao Wang,\authormark{1,2} Chunxu Zhao,\authormark{3} Guangquan Wang,\authormark{3} Ruichun Wang,\authormark{4} Honghai Wang,\authormark{4} Fei zhou,\authormark{5} Jimin Nie,\authormark{6} Qing Chen,\authormark{7} Yong Zhao,\authormark{7} Qiang Zhang,\authormark{1,2} Jun Zhang,\authormark{1,2} Teng-Yun Chen,\authormark{1,2,*} and Jian-Wei Pan\authormark{1,2}}

\address{\authormark{1}Hefei National Laboratory for Physical Sciences at Microscale and Department of Modern Physics, University of Science and Technology of China, Hefei, Anhui 230026, China\\
\authormark{2}CAS Center for Excellence and Synergetic Innovation Center in Quantum Information and Quantum Physics, University of Science and Technology of China, Hefei, Anhui 230026, China\\
\authormark{3}Network Technology Research Institute, China Unicom Network Communications Corporation Limited, Beijing 100048, China\\
\authormark{4}State Key Laboratory of Optical Fiber and Cable Manufacture Technology, Yangtze Optical Fibre and Cable Joint Stock Limited Company, Wuhan, Hubei 430073, China\\
\authormark{5} Jinan Institute of Quantum Technology, Shandong Academy of Information Technology, Jinan, Shandong 250101, China \\
\authormark{6}CAS Quantum Network Corporation Limited, Shanghai  201315, China\\
\authormark{7}QuantumCTek Corporation Limited, Hefei, Anhui 230088, China}

\email{\authormark{*}tychen@ustc.edu.cn} 



\begin{abstract}
Quantum key distribution (QKD) provides information-theoretic security based on the laws of quantum mechanics. The desire to reduce costs and increase robustness in real-world applications has motivated the study of coexistence between QKD and intense classical data traffic in a single fiber. Previous works on coexistence in metropolitan areas have used wavelength-division multiplexing, however, coexistence in backbone fiber networks remains a great experimental challenge, as Tbps data of up to 20 dBm optical power is transferred, and much more noise is generated for QKD. Here we present for the first time, to the best of our knowledge, the integration of QKD with a commercial backbone network of 3.6 Tbps classical data at 21 dBm launch power over 66 km fiber. With 20 GHz pass-band filtering and large effective core area fibers, real-time secure key rates can reach $4.5$ kbps and $5.1$ kbps for co-propagation and counter-propagation at the maximum launch power, respectively. This demonstrates feasibility and represents an important step towards building a quantum network that coexists with the current backbone fiber infrastructure of classical communications.
\end{abstract}



\section{Introduction}

Quantum key distribution (QKD) \cite{RMP2002gisinQKD}, based on the laws of quantum physics, provides proven unconditional security for data communication between remote users. Since its introduction in $1984$ \cite{BB84}, experimental efforts have been made to achieve long distance and high key rate in optical fibers, such as  the recent record demonstration of 404 km \cite{yin2016mdi404} and Mbits per second (Mbps) secure key generation \cite{dynes2016ultra}. So far, QKD has become one of the most mature and advanced quantum information technologies ready for use. To demonstrate its reliability and robustness over long periods of time, field experiments of QKD based secure communications for metropolitan networks and intercity links have been performed  \cite{hughes2000quantum,elliott2005current,chen2009field,peev2009secoqc,sasaki2011field,stucki2011long,qiu2014quantum,wang2014field,tang2016measurement}.

However, in order to protect ultra-weak QKD signals, all of the above achievements were performed in dark fibers with a wavelength set around $1550$ nm. This implies dedicated fiber installations for quantum communication networks, which bears cost penalties in fiber leasing and maintenance, as well as limitations on the network scale. In fact, to reduce cost and increase fiber transmission efficiency, classical communications (CC) have already exploited methods such as time-division multiplexing (TDM) and wavelength-division multiplexing (WDM) to achieve data throughput of up to Terabits per second (Tbps) and ultra-long transmission distances \cite{essiambre2010capacity}. Therefore, it is highly desired to integrate QKD with CC in existing fiber infrastructures and to expand the scalability of QKD networks.

The scheme of simultaneously transmitting QKD with conventional data was first introduced by Townsend in $1997$ \cite{townsend1997simultaneous}. Using coarse wavelength-division multiplexing (CWDM), a QKD channel at $1300$ nm was multiplexed with a conventional $1.2$ Gbps data channel at $1550$ nm over $28$ km installed fiber. Although no privacy amplification or yield of secret key was reported \cite{townsend1997simultaneous}, it provided a blueprint for the coexistence works that followed. A series of QKD experiments integrating with various classical channels have been demonstrated \cite{chapuran2009WDMqkd,choi2010PONqkd,eraerds2010WDMqkd,patel2012WDMqkd,patel2014DWDMqkd,wang2015WDMqkd,kumar2015coexistence,huang2015continuous,frohlich2015PONqkd,dynes2016ultra}, and perspectives on the obstacles and approaches to share existing fiber infrastructures among quantum and classical channels have been discussed \cite{1253165,subacius2005backscattering,qi2010feasibility,aleksic2015qkdWDMperspectives}.

Currently, by using spectral and temporal controls, state-of-the-art developments have been made to realize co-propagation of QKD with one $100$ Gbps dense wavelength-division multiplexing (DWDM) data channel in $150$ km ultra-low loss fiber at $-5$ dBm launch power \cite{frohlich2017long}. By setting QKD wavelength to $1310$ nm and inserting DWDM filters with bandwidth of $100$ GHz before QKD receivers to improve out-of-band noise rejection, co-propagation of QKD with classical traffic was recently demonstrated over $80$ km fiber spools \cite{wang2017long}. In that experiment, by assembling the CC system with high-performance affiliated equipment in the laboratory, data transmission at Tbps level was achieved at $11$ dBm launch power. A field trial of simultaneous QKD transmission and four $10$ Gbps encrypted data channels was implemented over $26$ km installed fiber at $-10$ dBm launch power \cite{choi2014fieldWDMqkd}.

However, the integration of QKD with existing backbone network is different from the previous implementations. On the one hand, backbone networks use complex classical communication systems with extremely high-level stability, reliability, and redundancy. Therefore, in realistic backbone networks, launch power for Tbps level classical data reaches $\sim20$ dBm, which results in significantly stronger Raman scattering to QKD than in Ref. \cite{wang2017long}. Using the previous 100 GHz pass-band filtering, key generation will be difficult, and coexistence may not be achieved in realistic backbone networks. On the other hand, such field integration may suffer from environmental factors and social activities that are not encountered in laboratory demonstrations. To close the gap between laboratory demonstrations and practical applications, there are many great technical challenges to overcome for the integration of QKD with backbone fiber infrastructures.

In this work, we present for the first time, to the best of our knowledge, the coexistence of QKD and commercial backbone network of $3.6$ Tbps classical data over $66$ km fiber at the maximum launch power, i.e., $21$ dBm. With a $20$ GHz pass-band filter, we achieve co-propagation of QKD and 21 dBm classical data over standard single-mode fiber with $3.0$ kbps secure key rate and $2.5\%$ quantum bit error rate (QBER). Using low-loss fiber with large effective core area, secure key rates can be further increased to $4.5$ kbps and $5.1$ kbps for co-propagation and counter-propagation, respectively. Contributions of filter bandwidth, fiber loss, and fiber effective core area to the quantum signal-to-noise ratio (QSNR) and secure key rate are modeled and analyzed. The QKD stability and the coexistence effects on CC are also investigated.

\section{Experiment}

\begin{figure}[h]
\centering
{\includegraphics[width=\linewidth]{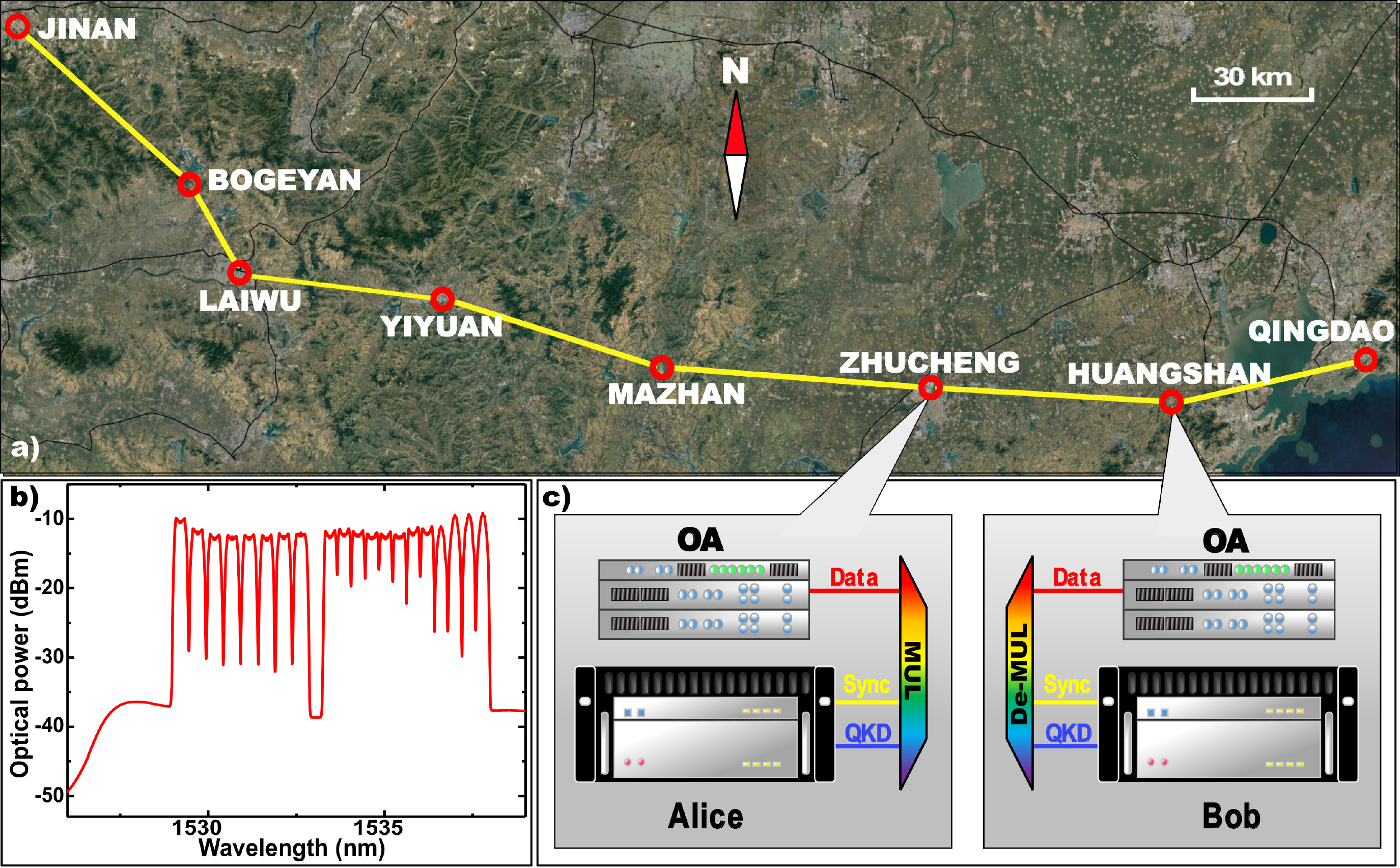}}
\caption{a) Field configurations of the experiment. The single yellow line represents the deployed two fiber links between communication stations. b) Spectrum of the CC measured at Zhucheng. The $200$ Gbps channels are shown as the root-raised cosine waveforms, while the $100$ Gbps channels are shown as the Gaussian waveforms. c) Coexistence schematics of the experiment. The wavelengths of QKD, sync, and data signals are $1310$ nm, $1490$ nm, $1528\sim1538$ nm, respectively.}
\label{fig:fig1}
\end{figure}

The field experiment is performed over a backbone loop network with a total installed fiber distance of $800$ km and eight nodes spanning across Shandong province, China, deployed by China United Telecom (China Unicom), as shown in Fig. \ref{fig:fig1}a. Considering fiber loss between the nodes and QKD system performance, we assign Alice to node Zhucheng ($119^{\ensuremath{^\circ}}24^{\prime}43^{\prime\prime}$ N $36^{\ensuremath{^\circ}}3^{\prime} 00^{\prime\prime}$ E) and Bob to node Huangshan ($119^{\ensuremath{^\circ}}59^{\prime}50^{\prime\prime}$ N, $36^{\ensuremath{^\circ}}2^{\prime}4^{\prime\prime}$ E), which has in-between $66$ km installed fiber deployed along highways, bridges, and tunnels. Meanwhile, the transmitter and receiver  of CC are placed in the Jinan laboratory for performance analysis, while optical amplifiers (OA) are located in other stations, controlled by internet monitors for optical gain adjustments. We verify the performances of both QKD system and CC system on three types of fibers with different effective core areas (whereas the fibers connecting other nodes are all G654 fibers with an effective core area of $110$ $\mu m^2$), whose specifications are listed in Table \ref{tab:table1}, with fiber links G652-1, G654-110-1, G654-130-1 for co-propagation and fiber links G652-2, G654-110-2, G654-130-2 for counter-propagation. All fibers on the backbone network follow the International Telecommunication Union standards. Since these G654 fibers possess larger effective core areas than the standard single-mode fiber, it can be used to lower optical effects at the fundamental light-matter interaction level in coexistence experiments and thus to enhance QKD performances.

\begin{table}[ht!]
\centering
\caption{Field fibers specifications}
\begin{tabular}{cccccc}
\hline
\multirow{2}{*}{Fiber type} & \multicolumn{2}{c}{Att (dB/km)}  & \multicolumn{2}{c}{Total loss (dB)} & Aeff \\ 
 & $1550$nm & $1310$nm & $1550$nm & $1310$nm & $\mu m^2$ \\
\hline
G652-1 & $0.197$ & $0.337$ & $13.01$ & $22.21$ & $80$ \\
G652-2 & $0.196$ & $0.338$ & $12.94$ & $22.29$ & $80$ \\ 
G654-110-1 & $0.184$ & $0.300$ & $12.13$ & $19.83$ & $110$ \\ 
G654-110-2 & $0.174$ & $0.288$ & $11.51$ & $19.03$ & $110$ \\ 
G654-130-1 & $0.210$ & $0.347$ & $13.84$ & $22.88$ & $130$ \\ 
G654-130-2 & $0.208$ & $0.348$ & $13.73$ & $22.95$ & $130$ \\
\hline
\end{tabular}
  \label{tab:table1}
\end{table}

For classical data traffic, we adopt a commercial hybrid system from China Unicom (Alcatel-Lucent 1830PSS-64) with aggregate transmission rate of $3.6$ Tbps, which is achieved by four line cards with $50$ GHz bandwidth and four line cards with $62.5$ GHz bandwidth. In each line card, the super-channel scheme is used, consisting of two sub-carriers that carry $200$ Gbps polarization division multiplexing with $8$-quadrature amplitude modulation format. In addition, four $100$ Gbps line cards with quadrature phase shift keying modulation format are included in the CC system. The CC spectrum is shown in Fig. \ref{fig:fig1}b, including $20$ channels covering about $1528\sim1538$ nm. The total launch power can be tuned from $8$ to $21$ dBm, with optimal launch power is $\sim18$ dBm. Utilizing pseudo-random binary sequence generated by a test instrument to serve as the classical data throughout the experiment and varying the launch power, the parameters of Q-factor and OSNR are measured. When the launch power is reduced down to a low level, e.g., $11$ dBm as in \cite{wang2017long}, the bit error rate (BER) is close to the bound for forward error correction, which results in the failure of the data transmission of CC.

For QKD, we develop a polarization encoding, decoy-state BB84 protocol-based system. Signal pulses at $1310$ nm are prepared at Alice\textquotesingle s site with a repetition rate of $625$ MHz. The pulse train is internally modulated to $70$ ps width and further shaped to $1310 \pm 0.036$ nm at full width at half maximum (FWHM) with a $10$ GHz fiber Bragg grating (FBG), which is used to compensate dispersion effects and to guarantee pulse registration within the detector effective gating width. The mean photon numbers of the signal state $\mu$, decoy state $\nu$, and vacuum state $\omega$ are $0.6$, $0.2$, $0$ \cite{wang2007laserdecoy}, respectively, with an emission ratio of $P_{\mu} : P_{\nu} : P_{\omega} = 6 : 1 : 1$, which are controlled by a physical random number generator implemented in a field programmable gate array (FPGA). Using two polarization beam splitters and a polarization controller, four nonorthogonal states are generated with output powers adjusted to single-photon level using a variable optical attenuator. The received photons at Bob\textquotesingle s site are recorded by four InGaAs/InP single-photon detectors (SPDs) \cite{liang2012fully,zhang2015advances}, which operated at $1.25$ GHz gating frequency with a detection efficiency of $11\%$ and a dark count rate per gate of $3\times10^{-7}$ in average. The detector effective gating width is set to ${\sim}180$ ps to provide temporal filtering while maintaining maximum detection efficiency, and the detector dead time ($t_{\textrm{dead}}$) is $1$ $\mu s$. Automatic polarization feedback system is implemented using two electric polarization controllers at Bob's site to adjust the extinction ratio between the orthogonal states \cite{chen2010metropolitan}. Visibility reaches $\sim21.5$ dB after feedback control, corresponding to a $0.7\%$ optical misalignment error rate ($e_{d}$). Meanwhile, synchronized clock signal between Alice and Bob (Sync) at $1490$ nm propagates along QKD signals with a repetition rate of $100$ kHz and a received optical power of $-53$ dBm.

After the pulses are registered at Bob\textquotesingle s detectors, a series of post-processing for real-time secure key extraction is performed. First, the public discussion channel between Alice and Bob is established through pre-stored QKD keys, which are periodically replaced with fresh keys. The keys are fed into Toeplitz Hash algorithms using linear feedback shift register to implement authentication \cite{krawczyk1994lfsr}. Next, key sifting and basis sifting are executed, followed by error correction (EC) using Winnow codes with a correction efficiency of $1.2{\sim}1.5$, and error verification (EV) using CRC-64 algorithm. Afterwards, privacy amplification (PA) is performed to eliminate information leakage to Eve during EC and EV. An estimated final key rate is calculated using standard decoy-state method \cite{wang2005beating,lo2005decoy,ma2005practical,wang2016tight} to determine a PA factor, defined as the ratio of the estimated final key rate to the corrected key rate after EV. With the PA factor, an exact Toeplitz matrix is constructed to extract final secure keys from the corrected keys. Considering the real-time key extraction period, the block size for parameter estimation is set to $500$ kbits to guarantee the freshness of final keys. Also, considering statistical fluctuations, $7$ standard deviations are used to guarantee the security of final keys. Further, by evaluating QBER using the decoy-state method, a strict upper bound is set to $4\%$ for redundancy control, which corresponds to a QSNR lower bound of ${\sim}11$ dB, to guarantee stable key generation. 

To achieve coexistence, WDM filters are applied, forming a multiplexing module (MUL) at Alice and a de-multiplexing module (De-MUL) at Bob, see Fig. \ref{fig:fig1}c. The insertion losses at $1550$ nm and $1310$ nm are respectively $0.86$ dB and $0.30$ dB for MUL, while $0.87$ dB and $2.50$ dB for De-MUL. The isolation of $1310$ nm from $1550$ nm at MUL is $50$ dB, while that of $1550$ nm from $1310$ nm at De-MUL is $120$ dB. Combining with a circulator, a temperature-compensated FBG with a reflection band of $1310 \pm 0.087$ nm at FWHM forms a $20$ GHz filter in De-MUL. The crosstalk between QKD and CC channels is measured, with a noise level of $\sim60$ counts per second (cps), which is much lower than the dark count rates of the SPDs and thus negligible in the experiment.

When QKD is integrated with CC, the strong CC signals may cause severe impairments on the weak QKD signals through fundamental light-matter interactions. Possible sources of noise in the QKD wavelength from CC channels include four-wave mixing, Brillouin scattering, and Raman scattering. By allocating the QKD wavelength to $1310$ nm, a $\sim200$ nm gap is set between the QKD and CC signals, so the noise photons caused by four-wave mixing in the QKD channel is negligible. In addition, because of its low bandwidth ($\sim10$ GHz), Brillouin scattering from the CC channels do not affect QKD. In contrast, Raman scattering lead to large spectral shifts from the CC channel wavelengths, with an intensity maximum at a shift of $\sim13$ THz and maximum offset parameter to more than $50$ THz \cite{aleksic2015qkdWDMperspectives}. Stimulated Raman scattering has a threshold of $\sim600$ mW for $20$ km single-mode fiber at $50$ $\mu m^2$ effective core area \cite{agrawal2007nonlinear}, so considering the fiber length and CC launch powers of our experiment, this effect from CC channels in the QKD channel is negligible. Therefore, in this experiment, the photons generated by spontaneous Raman scattering (SRS) from the CC channels is the only optical scattering noise in the QKD channel. Furthermore, with the $20$ GHz filter, SRS photons can be further reduced and hence the QSNR is enhanced, which suggests the coexistence possibility with CC at the maximum launch power, despite the fact that the $20$ GHz filter has an insertion loss of $1.9$ dB compared to that of $0.5$ dB for $100$ GHz filter in Ref. \cite{wang2017long}.

\section{Results and discussion}

\begin{figure}[h!]
\centering
  \includegraphics[width=\linewidth]{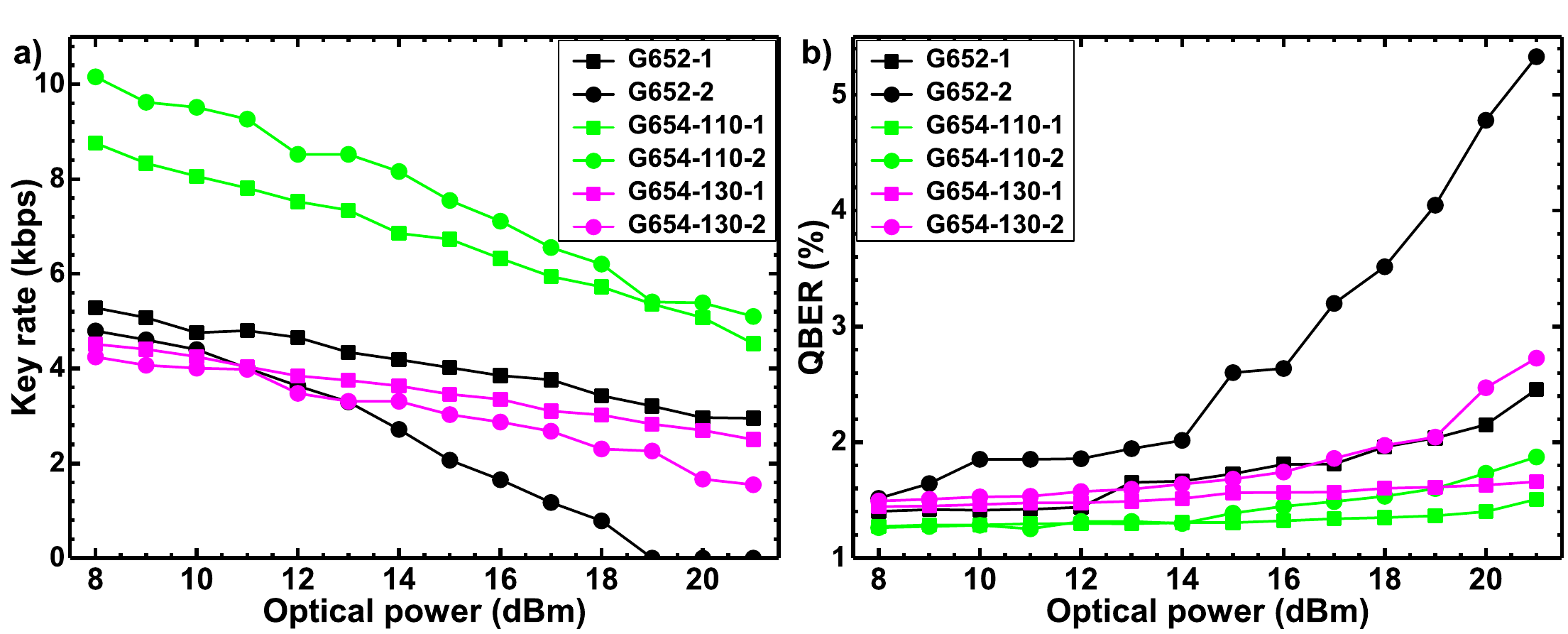}\\
  \caption{Secure key rate (a) and QBER (b) of QKD as a function of CC optical power. Solid squares: co-propagation; solid circles: counter-propagation. }
\label{fig:fig2}
\end{figure}

By adjusting the launch power of CC from $8$ to $21$ dBm, we measure the secure key rate and QBER of QKD, as shown in Figs. \ref{fig:fig2}a and b. For G652 fiber, $3.0$ kbps 
secure key rate is achieved at the maximum launch power for co-propagation. For counter-propagation, secure key rates significantly drop given the launch power higher than $18$ dBm. This is due to the fact that much more SRS photons are created than in the case of co-propagation \cite{chapuran2009WDMqkd,choi2010PONqkd,subacius2005backscattering}, corresponding to a lower QSNR, given the same length of fiber and launch power. At $19$ dBm, QBER reaches $4.0\%$ and no secure keys are generated, as shown in Figs. \ref{fig:fig2}a and b. These results indicate the possibility of coexistence between QKD 
and classical backbone network in standard single-mode fiber, both for co-propagation at the maximum launch power and for counter-propagation by slightly decreasing the launch power.

\begin{figure}[h!]
\centering
  \includegraphics[width=9.5cm]{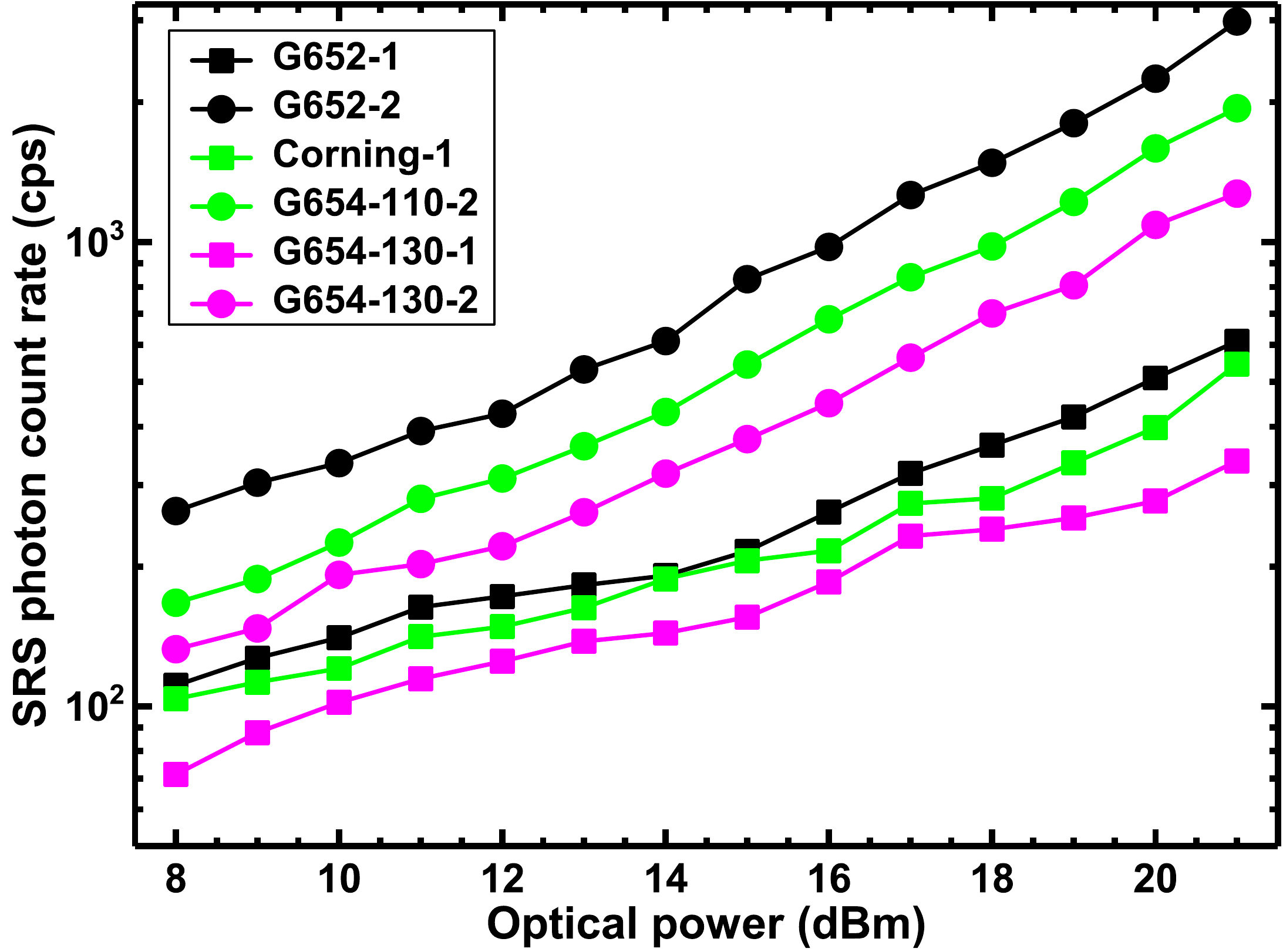}\\
  \caption{Photon count rate generated by SRS. By sending classical signals through the $66$ km field fiber along either the same direction of QKD signals (co-propagation, solid squares) or the opposite direction (counter-propagation, solid circles), photons are recorded by the SPDs. During this process, no QKD signals are transmitted. At each point, four SPD count rates are averaged. For co-propagation and counter-propagation, G654-130 fiber exhibits the lowest SRS photon count rates, with $\sim 60\%$ reduction compared with the G652 fiber.}
\label{fig:fig3}
\end{figure}

The best QKD performance is achieved in G654-110 fiber. At the maximum launch power, key rates reach $4.5$ kbps and $5.1$ kbps for co-propagation and counter-propagation, respectively. The higher key rate in counter-propagation is due to the relatively lower fiber attenuation of G654-110-2 fiber, see Table \ref{tab:table1}. For G654-130 fiber, the created SRS photons are much less than that in G652 fiber, as shown in Fig. \ref{fig:fig3}. However, due to the characteristic of higher fiber attenuation, the key rate for co-propagation is lower than G652 fiber. For counter-propagation, given a launch power larger than $13$ dBm, contributions due to SRS in G652 fiber is much more significant than fiber attenuations, which results in a lower QSNR than G654-130 fiber.

Further, we model the key rates using different fibers. Considering fiber attenuation and the SRS of different effective core areas, the QSNR is defined as
\begin{equation*}
\textrm{QSNR}=10 {log}_{10}(\frac{N_{\textrm{actual}}  (1-P_{\textrm{after}}-e_{d})}{N_{\textrm{SRS}}+N_{\textrm{dark}}+N_{\textrm{after}}}),
\label{eq:QSNR}
\end{equation*}
where $N_{\textrm{actual}}$ is the actual QKD signal count rate, $N_{\textrm{SRS}}$ is the SRS photon count rate, $N_{\textrm{dark}}$ is the detector dark count rate, $P_{\textrm{after}}$ is the detector afterpulse probability  ($\sim0.5\%$ in our experiment), and $N_{\textrm{after}}$ is the afterpulse count rate, calculated by $N_{\textrm{after}}=N_{\textrm{actual}}\cdot P_{\textrm{after}}$. Due to the detector dead time, $N_{\textrm{actual}}={\nicefrac{N_{\mu}}{(1+\frac{N_{\mu} \cdot t_{\textrm{dead}}}{4})}}$, where $N_{\mu}$ is the QKD signal count rate that would be obtained if dead time was negligible. $N_{\mu}$ and $N_{\textrm{SRS}}$ are calculated, while $N_{\textrm{actual}}$, $N_{\textrm{dark}}$, and $t_{\textrm{dead}}$ are measured.

\begin{table*}[h!]
\centering
\caption{Simulation results of QSNR and secure key rate in different cases. The attenuation coefficients of standard fiber and low-loss fiber are 0.337 dB/km at 1310 nm and 0.197 dB/km at 1550 nm, and 0.288 dB/km at 1310 nm, 0.174 dB/km at 1550 nm, respectively.}
\begin{tabular}{ccccc}
\hline
\multirow{2}{*}{Aeff ($\mu m^2$)} & \multicolumn{2}{c}{Co-propagation}  & \multicolumn{2}{c}{Counter-propagation} \\
  & {QSNR (dB)} & {Key rate (kbps)} & {QSNR (dB)} & {Key rate (kbps)} \\
\hline
\multicolumn{5}{c}{20 GHz filtering, low-loss fiber}\\
\hline
80 & $50.5$ & $5.5$ & $18.3$ & $3.2$\\
110 & $65.8$ & $5.9$ & $29.5$ & $4.9$ \\
130 & $71.1$ & $6.0$ & $34.5$ & $5.2$ \\
\hline
\multicolumn{5}{c}{20 GHz filtering, standard fiber}\\
\hline
80 & $36$ & $2.2$ & $10.0$ & $-$\\
110 & $45.8$ & $2.3$ & $16.6$ & $1.2$ \\
130 & $49.2$ & $2.4$ & $19.8$ & $1.5$ \\
\hline
\multicolumn{5}{c}{100 GHz filtering, low-loss fiber}\\
\hline
80 & $14.0$ & $1.8$ & $2.8$ & $-$\\
110 & $23.5$ & $3.6$ & $5.8$ & $-$ \\
130 & $28.1$ & $4.1$ & $7.5$ & $-$ \\
\hline
\multicolumn{5}{c}{100 GHz filtering, standard fiber}\\
\hline
80 & $10.4$ & $-$ & $1.0$ & $-$\\
110 & $17.5$ & $1.1$ & $2.7$ & $-$ \\
130 & $20.8$ & $1.4$ & $3.6$ & $-$ \\
\hline
\end{tabular}
  \label{tab:table2}
\end{table*}

We then calculate $N_{\textrm{SRS}}$, following the methods in Refs.  \cite{chapuran2009WDMqkd,choi2010PONqkd} to obtain the SRS coefficient ($\beta_{20\, \textrm{GHz}}$) of the three fibers. By averaging the results in Fig. \ref{fig:fig3} with different launch powers, for both co-propagation and counter-propagation, we calculate the values of $\beta_{20\, \textrm{GHz}}$ as $18$, $10$, and $8$ $\nicefrac{\textrm{cps}}{\textrm{dBm}\cdot \textrm{km}}$ for G652, G654-110, and G654-130, respectively.

Following the above model, we evaluate the contributions of the major factors, i.e., filter bandwidth, fiber attenuation, and fiber effective core area, to the key rate of QKD coexisting with $21$ dBm launch power over $66$ km transmission distance. The simulated results are shown in Table \ref{tab:table2}, from which one can find out three facts. First, narrow pass-band filtering in such coexistence experiment is necessary. As shown in Table \ref{tab:table2}, using $100$ GHz filtering, the QSNR values for counter-propagation in three fibers are below $8$ dB, without generating secure keys. In particular, no keys are generated for both co-propagation and counter-propagation in standard fiber with $100$ GHz filtering. This means QKD cannot be integrated with backbone network using the same coexistence scheme of Ref. \cite{wang2017long}. Second, using low-loss fiber, the QSNR is enhanced by $15.4$ dB in average, and the key rate can be increased twice at least. Third, since large effective core area reduces the optical power density of the CC, the QSNR can be improved, which brings moderate increase in key rate. By combining $20$ GHz filtering, low-loss fiber with $130$ $\mu m^2$ effective core area, the highest secure key rates of QKD for co-propagation and counter-propagation may reach $6.0$ kbps and $5.2$ kbps, respectively.

\begin{figure}[h!]
\centering
\includegraphics[width=9.5cm]{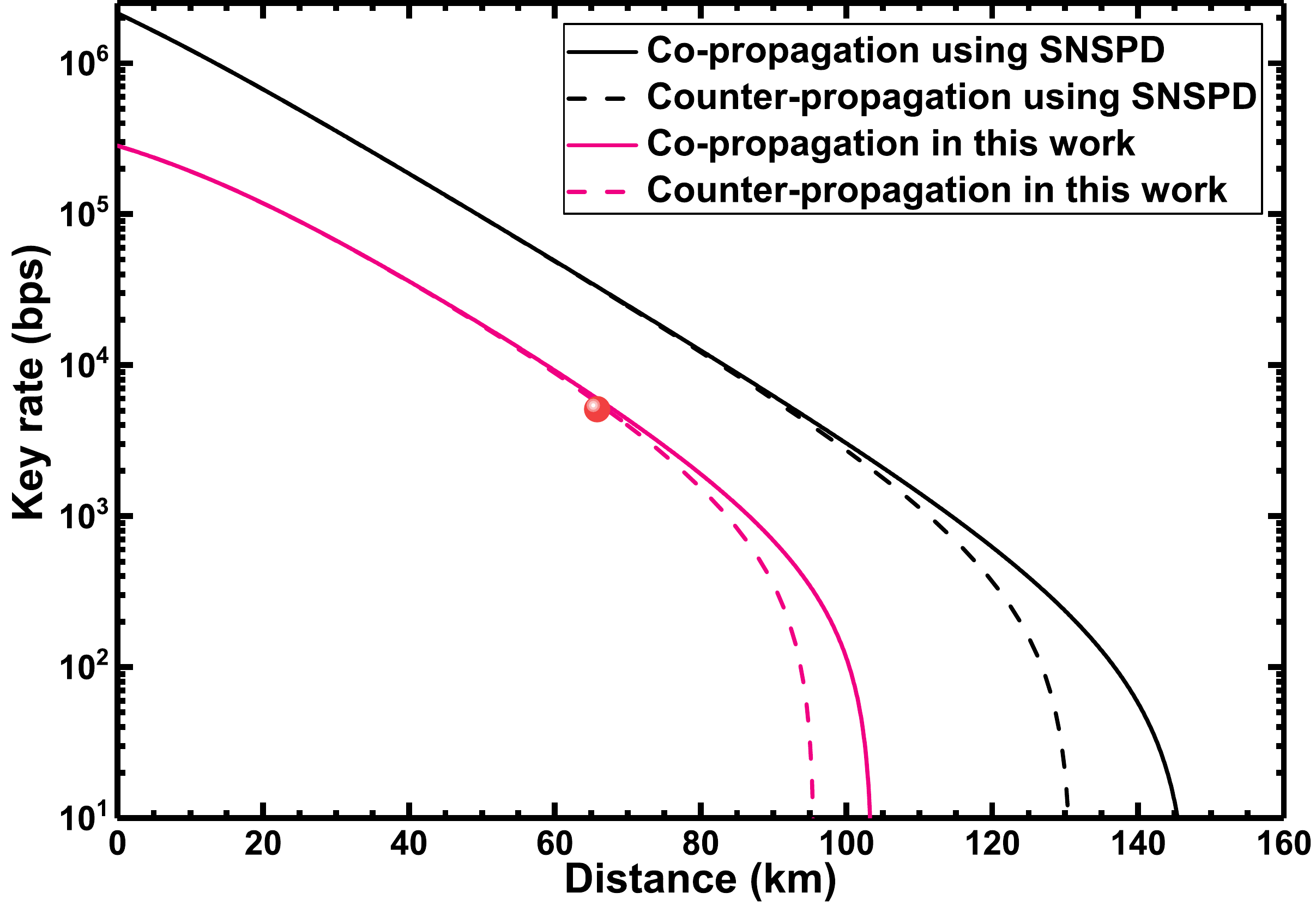}
  \caption{Calculated secure key rates of QKD as a function of distance for co-propagation (solid lines) and counter-propagation (dashed lines) with $21$ dBm launch power in G654-110-2 fiber. The red lines represent the results with the QKD system in the experiment, while the black lines represent the results using SNSPDs instead of InGaAs/InP SPDs for the QKD system. The solid red dot represents the measured result in the experiment.}
\label{fig:fig4}
\end{figure}
In the experiment, we have demonstrated integrating QKD with $3.6$ Tbps classical data over $66$ km installed fiber. Further results for longer transmission distances using G654-110-2 fiber at the maximum launch power are simulated, as shown in Fig. \ref{fig:fig4}. At $70$ km and $80$ km, secure key rates are respectively $4.4$ kbps and $1.9$ kbps for co-propagation , and $4.0$ kbps and $1.5$ kbps for counter-propagation. Even at 90 km, secure key rate may still reach $0.7$ kbps for co-propagation. This suggests that even integrating with CC at a typical span distance of $80$ to $100$ km secure key rate of QKD is still sufficient for one-time pad voice and text encryptions. Using superconducting nanowire single-photon detectors (SNSPDs) instead of InGaAs/InP SPDs used in the experiment, the secure key rate can be substantially increased. We perform the theoretical calculations of key rate for co-propagation and counter-propagation at $21$ dBm launch power in G654-110-2 fiber, using the calibrated parameters of the SNSPDs in our laboratory \cite{you2014SNSPD}, i.e., $45\%$ detection efficiency at $1310$ nm and $30$ cps dark count rate. The key rate is calculated following the decoy-state approach in Ref. \cite{ma2005practical}. As shown in Fig. \ref{fig:fig4}, when the SNSPDs are used, the key rates for co-propagation and counter-propagation can be significantly increased to $32.7$ kbps and $32.3$ kbps at $66$ km, respectively, while the maximum coexistence distances reach $145$ km and $130$ km for co-propagation and counter-propagation, respectively.
\begin{figure}[h!]
\centering
  \includegraphics[width=9.5cm]{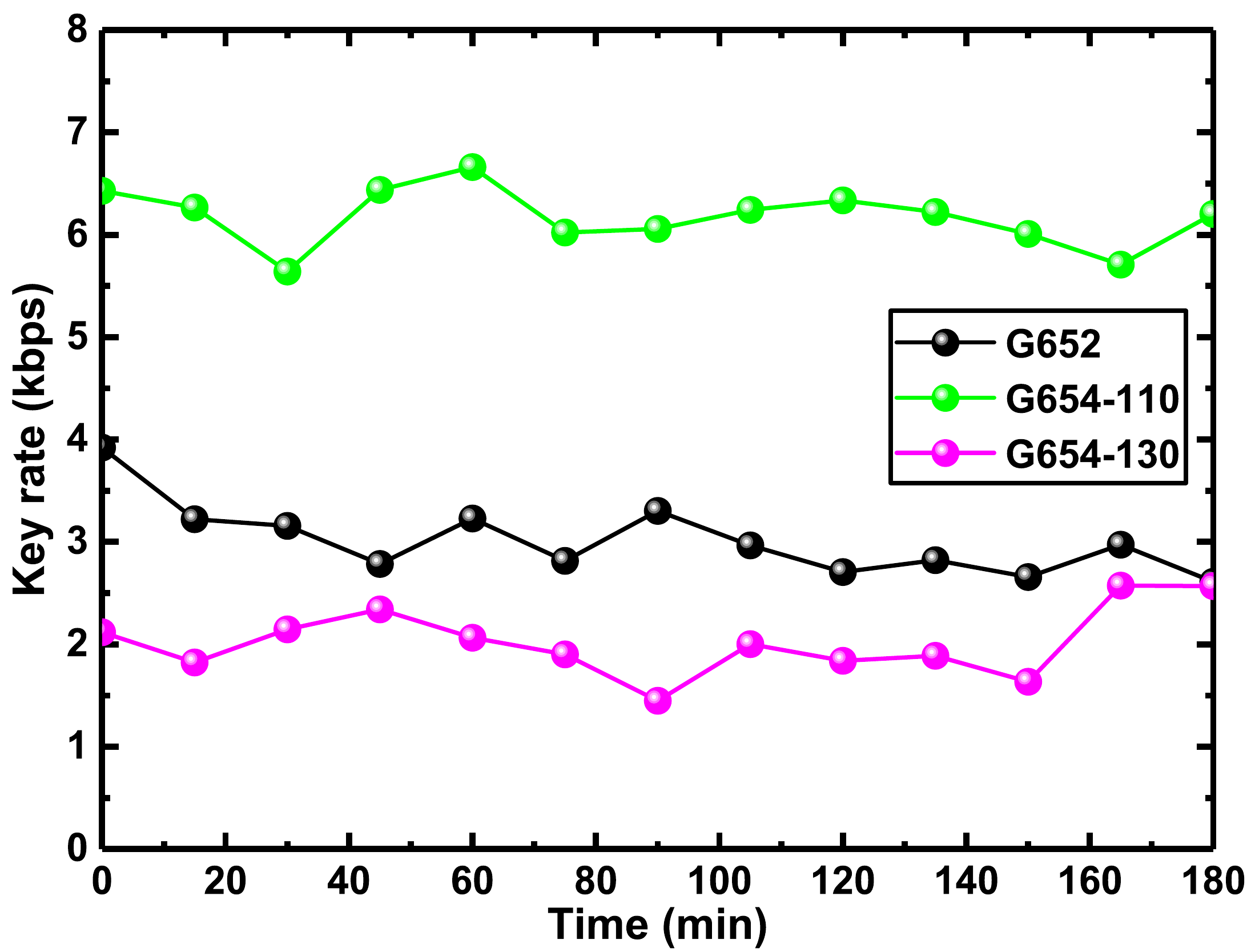}\\
  \caption{The stability test of secure key rate in three fibers.}
\label{fig:fig5}
\end{figure}

We further investigate the stability of QKD coexisting with CC systems. Due to the limited usage time in the backbone communication stations, we perform co-propagation measurement in three fibers for $3$ hours each, with $\sim18$ dBm launch power. The average key rates are $6.2$ kbps, $3.0$ kbps, and $2.0$ kbps for G654-110, G652, and G654-130 fibers, respectively, as shown in Fig. \ref{fig:fig5}. Despite the effects from social activities such as traffic, electricity, and constructions, the stability test results indicate that our QKD systems can be reliably integrated with the CC systems of backbone networks.

\begin{figure}[ht!]
\centering
\includegraphics[width=\linewidth]{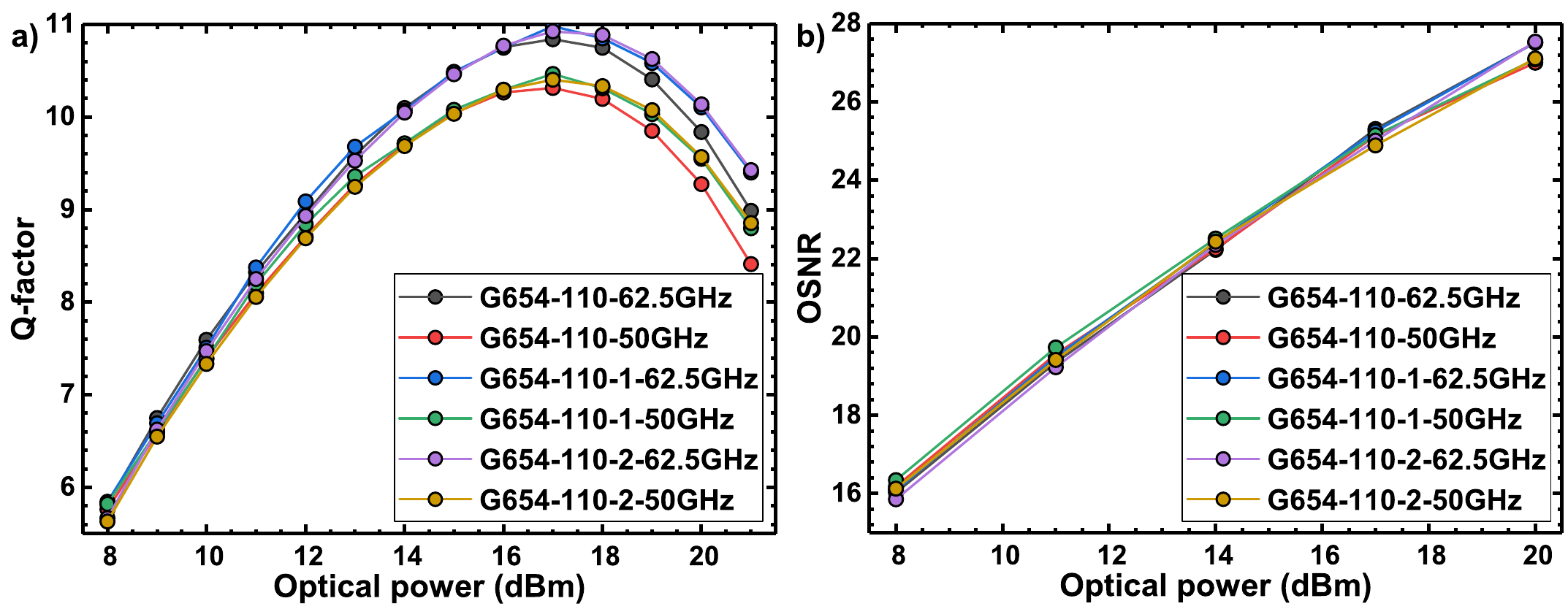}
  \caption{Q-factor (a) and OSNR (b) of CC system as the function of optical power in G654-110 fiber by applying $62.5$ GHz and $50$ GHz bandwidth carriers, respectively. G654-110: no co-existence; G654-110-1: co-propagation; G654-110-2: counter-propagation. }
\label{fig:fig6}
\end{figure}

Finally, we measure the CC performance changes in the presence of QKD. The measured results in G654-110 fiber are plotted in Figs. \ref{fig:fig6}a and b, which show that the inclusion of QKD brings negligible variations on OSNR of CC system. Nevertheless, due to insertion losses of WDM modules, CC launch power is lowered and the optical effects between adjacent CC channels during data transmission are reduced, which induces a slight increase of Q-factor, particularly with launch powers higher than $16$ dBm, as shown in Fig. \ref{fig:fig6}a.

\section{Conclusion}

In summary, we have achieved for the first time, to the best of our knowledge, the coexistence of QKD with a commercial backbone network of 3.6 Tbps classical data at the maximum launch power over 66 km fiber. The key factors affecting the coexistence performance such as filter bandwidth, fiber attenuation, and fiber effective core area are modeled and analyzed. Our work validates the feasibility to build a quantum network coexisting with current backbone fiber infrastructures of classical communications. Utilizing superconducting nanowire single-photon detectors and increasing the repetition rate of the QKD system, the secure key rate and coexistence distance can be further improved. Meanwhile, in the presence of untrusted network nodes and detection attacks, measurement-device-independent QKD \cite{tang2016measurement} provides an effective approach and its coexistence with CC deserves future investigations. Using silicon photonics technologies, low-cost, chip-based QKD devices \cite{sibson2017chip} may be fully integrated inside CC modules, offering wide accessibility and large-scale application. 

In this work, the effects of transient problems caused by the dropping or adding data channels are not included, and the spectral bandwidth of the classical channels does not cover all the C$+$L-bands. Such interesting topics deserve future investigations.

\section*{Funding}

National Key R\&D Program of China (2017YFA0303900, 2017YFA0304004); Anhui Provincial Natural Science Foundation (Grant No. 1508085J02); Priority R\&D Plan Project (2015GGX101035).

\section*{Acknowledgments}

We thank L.-J. Wang, Y.-L. Tang for useful discussions, and H. Yu for assistance in classical communication system tests. 

\end{document}